\documentstyle[12pt]{article}
\newcommand{\be}{\begin{eqnarray}}
\newcommand{\ee}{\end{eqnarray}}
\newcommand{\bdm}{\begin{displaymath}}
\newcommand{\edm}{\end{displaymath}}

\begin{document}
\begin{center}
{\LARGE {Mass hierarchy, mass gap and Corrections to Newton's \\
\vskip 0.3cm law on thick branes with Poincar\'{e} symmetry}}
\end{center}
\vskip 3mm
\begin{center}
{\bf \large {Nandinii Barbosa--Cendejas$^\dagger$$^\ddagger$\footnote{E-mail: nandinii@correo.fie.umich.mx},
Alfredo Herrera--Aguilar$^\dagger$$^*$$^\natural$\footnote{E-mail: aha@fis.unam.mx}, Konstantinos}}
\end{center}
\begin{center}
{\bf \large {Kanakoglou$^\natural$$^{**}$\footnote{E-mail: kanakoglou@hotmail.com}, Ulises Nucamendi$^\natural$\footnote{E-mail: ulises@ifm.umich.mx} and
Israel Quiros$^\star$\footnote{E-mail: iquiros6403@gmail.com}}}
\end{center}
\vskip 0.1cm
$^\dagger$Instituto de  Ciencias F\'{\i}sicas, Universidad Nacional Aut\'onoma de M\'exico, M\'exico.\\
$^\ddagger$Facultad de Ingenier\'{\i}a El\'ectrica, Universidad Michoacana de San Nicol\'as de Hidalgo, M\'exico.\\
$^*$Mesoamerican Centre for Theoretical Physics, Universidad Aut\'onoma de Chiapas,
M\'exico.\\
$^\natural$Instituto de F\'{\i}sica y Matem\'{a}ticas, Universidad Michoacana de San Nicol\'as de Hidalgo, M\'exico. \\
$^{**}$School of Mathematics, Faculty of Sciences, Aristotle University of Thessaloniki, Greece.\\
$^\star$Departamento de Matem\'aticas, CUCEI, Universidad de Guadalajara, M\'exico.

\begin{abstract}
We consider a scalar thick brane configuration arising in a 5D theory of gravity coupled to a
self--interacting scalar field in a Riemannian manifold. We start from known classical solutions
of the corresponding field equations and elaborate on the physics of the transverse traceless
modes of linear fluctuations of the classical background, which obey a Schr\"odinger--like
equation. We further consider two special cases in which this equation can be solved analytically
for any massive mode with $m^2\geq0$, in contrast with numerical approaches, allowing us to study
in closed form the massive spectrum of Kaluza--Klein (KK) excitations and to analytically compute 
the corrections 
to Newton's law in the thin brane limit. In the first case we consider a novel solution with
a mass gap in the spectrum of KK fluctuations with two bound states -- the massless 4D graviton
free of tachyonic instabilities and a massive KK excitation -- as well as a tower of continuous
massive KK modes which obey a Legendre equation. The mass gap is defined by the inverse of the
brane thickness, allowing us to get rid of the potentially dangerous multiplicity of arbitrarily
light KK modes. It is shown that due to this lucky circumstance, the solution of the mass
hierarchy problem is much simpler and transparent than in the thin Randall--Sundrum (RS)
two--brane configuration. In the second case we present a smooth version of the RS model with a
single massless bound state, which accounts for the 4D graviton, and a sector of continuous
fluctuation modes with no mass gap, which obey a confluent Heun equation in the Ince limit. (The
latter seems to have physical applications for the first time within braneworld models). For this
solution the mass hierarchy problem is solved with positive branes as in the Lykken--Randall (LR) 
model and the model
is completely free of naked singularities. We also show that the scalar--tensor system is stable
under scalar perturbations with no scalar modes localized on the braneworld configuration.
\end{abstract}



\section{Introduction}

Even if large extra dimensions -- in opposition to the KK picture where they are compact and small
-- have a spatial nature, they are quite similar to time in the sense that a thin $3$--brane has
only one point available along them. When considering time evolution of an object, it has no
access to past or future, it can exist just at one instant along the arrow of time.

The idea of living in an Universe with extra dimensions is quite old and has evolved in several
directions, depending on the problem under consideration: unification of gravity and
electromagnetism \cite{nordstrom,KK}, the cosmological constant problem \cite{rubakov}, dark
matter \cite{branedarkmatter}, the mass hierarchy problem \cite{masshierarchy,mgog}, localization
of gravity on thin \cite{rs}--\cite{Localgravity1}, and thick branes
\cite{dewolfe}--\cite{bazeiaetal}, etc. Moreover, localization of 4D gravity
\cite{ariasetal}--\cite{bhrs} as well as matter fields \cite{matterloc} on scalar thick brane
configurations was presented in the framework of Weyl geometries.

Within this context, an alternative mechanism to compactification and a solution to the mass
hierarchy problem were proposed in \cite{mgog}--\cite{lr} using a 5D world described by gravity, a
cosmological constant and a pair of thin branes. However, this setup leads to a singular 5D
manifold of orbifold type. Quite soon there have appeared several models that attempted to solve
this crucial problem from the gravitational point of view by smoothing out the brane configurations 
in several ways within high energy physics \cite{dewolfe}--\cite{bazeiaetal}, \cite{thickbranes} and 
also in cosmology \cite{kantietal1}--\cite{kantietal2}. Notwithstanding, the 5D 
spacetime of these thick braneworlds, usually generated by scalar fields, generically possesses naked 
singularities at the boundaries of the manifold or at $\pm\infty$ if a mass gap is present in the 
spectrum of KK massive excitations \cite{gremm2,bhrs,kantietal3}. These singularities 
can be made harmless for a brane observer by imposing unitary boundary conditions that guarantee 4D
energy and momentum conservation as in \cite{GMZ,cohenkaplan} (see also \cite{gremm2,dago}). By
analyzing the spectrum of KK fluctuation modes with imposed unitary boundary conditions, one
realizes that its continuum sector is projected out, leading to a theory in which just the
discrete part of the spectrum remains unitary. Thus, these KK massive modes die off rapidly enough
as we approach the naked singularity, a fact that provides a viable model from the physical point
of view. This picture is in agreement with the fact that the existence of a mass gap provides an
easy way to control the excitation of the KK gravitons and the corresponding energy loss into the
extra dimension as pointed out in \cite{bhrs} within the framework of thick braneworlds defined in a 
Weyl geometry. Thus, this suppression of delocalized KK gravitons
also renders harmless the presence of naked singularities at the boundaries of the 5D manifold. On
the other hand, it is worth mentioning that these naked singularities can be resolved either by
lifting the 5D geometry to higher dimensions or to an effective string/M theory since there are
some examples in the literature where 5D naked singularities actually correspond to non--singular
10D geometries \cite{warner}.

In this paper we continue previous analytical work done on localization of 4D gravity on thick
branes with Poincar\'{e} symmetry and we show that with the aid of a scalar field we can smooth
out the RS model, still providing a generalized compactification mechanism, solving the mass
hierarchy problem and analytically computing the corresponding corrections to Newton's law in a
stable braneworld configuration for two particular solutions that were independently obtained in
\cite{gremm1}--\cite{gremm2} and \cite{csakietal} and possess completely different physical properties:

\newpage

For one of them, called solution A) from now on, the mass hierarchy problem can be solved in a
much simpler and more transparent way than in the Randall--Sundrum, combined with the
Lykken--Randall thin brane approach \cite{rs,lr}. The novel feature of this mass hierarchy resolution 
is an analog quantum mechanical potential of linearized (around a 4D Poincar\'{e}--invariant 
background\footnote{It is worth noticing that this kind of mass gaps usually appear in the framework 
of braneworld models with a de Sitter 4D induced spacetime instead of a Minkowski 4D induced 
manifold.}) 5D gravity, that asymptotes to a positive constant value and leads to a mass gap in the 
KK--spectrum of the gravitational fluctuations, which obey a Legendre equation for arbitrary mass. 
Besides, the present thick brane approach represents a better alternative to solve the mass hierarchy 
than the RS two--brane model \cite{rs}, since the TeV brane can have positive tension (as in \cite{lr}),
contrary to the RS model. In this case the 5D manifold presents naked singularities at its
boundaries, however, by imposing unitary boundary conditions on the KK graviton spectrum one
obtains a 4D spacetime physically acceptable as in \cite{gremm2,dago}.

For another special solution, labeled by B), for which the graviton spectrum of KK massive
fluctuations obeys a confluent Heun equation (CHE) in the Ince limit (see \cite{ronveaux}--\cite{BF} 
for a review and some solutions to this equation) and
has no mass gap in the spectrum since the analog quantum mechanical potential asymptotically
vanishes, the mass hierarchy problem can be also solved in a similar way, leading to a smooth
realization of the LR model with positive branes, with the advantage that the 5D manifold is 
completely free of naked singularities. The confluent Heun
equation is also known as generalized spheroidal wave equation (GSWE). This equation, together
with the double confluent Heun equation (DCHE), have arisen in the context of several problems of
theoretical physics. However, it seems that so far the Ince limit of the confluent Heun equation
had no physical application \cite{figueiredo}. Thus, here we provide such a physical application
of this equation and its solutions for the first time within the framework of the braneworld
paradigm.

Therefore, while solution A) for our scalar--tensor model displays a theoretically and
phenomenologically interesting mass gap in the graviton spectrum of KK massive modes, naked
singularities are developed at the boundaries of the 5D manifold. This fact requires, in turn,
imposing unitary boundary conditions on the model in order to make the naked singularities
harmless. On the other hand, the solution B) yields a mass spectrum for the KK gravitons with no
mass gap between the relevant massless zero mode, the 4D graviton, and the massive KK excitations,
leading to a 5D spacetime free of naked singularities. These solutions complement each other in
some sense and provide a complete analysis of the considered scalar--tensor braneworld model.

Thus, we consider, first, a thick brane classical solution and subsequently make a suitable ansatz for 
its linear perturbations. Then, as customary, we recast the corresponding eigenvalue problem into the 
solution of a Schr\"odinger--like equation with an analog quantum mechanical potential $V_{QM}$ defined 
by the bulk curvature of the spacetime, i.e. by the warp factor. 
It turns out that for the two special solutions we are considering, this potential constitutes a 
modified P\"oschl--Teller potential for solution A) (like the one that appears in \cite{gremm2,bhrs}),
and a volcano potential with finite bottom for solution B) (as the potential that arises in \cite{bh3}
within the framework of a Weyl geometry). We then show that
4D gravity can be recovered by solving this equation for the massless zero mode in both cases. 
A qualitative (physical) analysis of the behavior of the massive KK--modes is performed by studying
the structure of the analog quantum mechanical potential of these solutions. For the first one,
solution A), we point out that there are two bound states: the massless one corresponding to a
localized 4D graviton free of tachyonic instabilities, and a massive excitation mode. These modes
are separated by a mass gap that is completely defined by an integration constant inversely
proportional to the thickness of the brane.\footnote{In \cite{bhrs} it was shown that this
constant is indeed proportional to the square root of the self--interacting coupling constant of
the scalar field within a purely geometric Weyl braneworld.} 
Starting from certain mass value, there 
is a continuum of massive KK fluctuations. As for the second solution, solution B), we see that there is a 
single massless bound state, the 4D graviton, and a continuum of massive excitations with no mass gap 
between them. Thus, in both cases there is a continuous spectrum of delocalized massive modes that
asymptote to plane waves.

We get more insight into the physics of these modes by analytically solving the corresponding
Schr\"odinger--like equations with arbitrary mass $m\ge 0$ along the extra dimension, and getting
explicit expressions for the massless and massive bound states, as well as for the continuous
tower of massive KK--excitations. The general solution in the first case, i.e. for solution A), is
written in terms of a linear combination of associated Legendre functions of first and second
kind, whereas, in the second case, for solution B), the relevant equation can be recast into the
Ince's limit of a confluent Heun equation \cite{ronveaux}--\cite{BF} whose solution can be
expressed as pairs of two--sided infinite series (in the sense that the summation index runs from
$-\infty$ to $\infty$) of hypergeometric functions and modified Bessel functions which possess an
arbitrary phase parameter that is introduced in order to ensure the convergence of these series.
In contrast to solutions to the original confluent and double confluent Heun equations which
asymptotically possess a normal Thom\'e behaviour, the solutions to the Ince limit of the CHE
behave at infinity as the so--called subnormal Thom\'e solutions (for details see, \cite{figueiredo,olver},
for instance), which, for our setup, correspond to plane waves, in complete agreement with the
physical picture of the Schr\"odinger--like problem. The solutions in each pair of series have the
same coefficients that satisfy three--term recurrence relations which, in turn, possess a dominant
and a minimal solution. Despite the mathematical difficulty in determining these coefficients, it
is possible to construct an explicit analytic minimal solution for them by implementing the
Miller's method \cite{gautschi} and fixing the arbitrary phase parameter (by solving the
corresponding characteristic equation). Thus, these explicit solutions allow us to analytically
study in closed form the behaviour of the massive modes in both cases, contrary to usual numerical
approaches (see, e.g. \cite{gremm1,bazeiaetal}) and also enables us to analytically compute the
corresponding corrections to Newton's law.

We then recall a quite general result given in \cite{Giovannini} which shows that the Schr\"odinger dynamical 
equations of the scalar fluctuations of the system can be expressed in a certain form inferred from 
supersymmetric quantum mechanics which guarantees that the corresponding spectrum of fluctuations is positive 
definite and there are no negative tachyonic modes, ensuring the stability of the brane configuration under the 
scalar sector of perturbations. We go further and show analytically that for the considered solutions of our scalar-tensor 
system, the corresponding quantum mechanical potentials are positive definite potential barriers, implying that 
there are no localized scalar perturbations in our braneworld model. 

By considering an scenario in which, in addition to the thick brane, a (in principle thin, positive tension) probe 
brane -- where the Standard Model (SM) particles are trapped -- is located at an appropriate distance away from the 
thick brane, we are able to solve the mass hierarchy problem with positive branes by following the procedure of 
References \cite{rs,lr} in both cases. Due to the existence of the mass gap in the graviton KK--spectrum of the first 
analyzed case, the solution is much simpler and transparent than in the Lykken--Randall thin brane configuration. 

A brief summary of the physical properties of both solutions and concluding remarks are finally given.

\section{Setup, solutions and their physical implications}

Consider the 5D Riemannian action given by \cite{dewolfe}--\cite{gremm2} (we take $M_{\ast}^3=1/8$ for the
time being)
\begin{equation}
\label{action} S_5=\int
d^5x\sqrt{|G|}\left[\frac{1}{4}R_5-\frac{1}{2}(\nabla\phi)^2-V(\phi)\right],
\end{equation}
where $\phi$ is a bulk scalar field and $V(\phi)$ is a self--interacting potential. We shall study
solutions which preserve 4D Poincar\'e invariance with the metric
\begin{equation}
\label{conflinee} {ds}_5^2=e^{2A(y)}\eta_{nm}dx^n dx^m+dy^2,
\end{equation}
where $e^{2A(y)}$ is the warp factor of the metric and depends just on the fifth dimension $y$;
$m,n=0,1,2,3$. We do not go into details about obtaining solutions to this system (see
\cite{dewolfe}--\cite{csakietal} for details), but just quote a pair of concrete solutions with different 
physical properties.

Let us first consider the following solution to the scalar--tensor model (\ref{action}) that was first obtained 
in \cite{gremm2} (this solution was also obtained in \cite{csakietal} in conformally flat 
coordinates):
\begin{eqnarray}
{\bf Solution \  A)}\qquad \label{sol} e^{2A}=\left\{\cos\left[a(y-y_0)\right]\right\}^b,\quad
\phi=\frac{\sqrt{3b}}{2}\ln\left\{\sec\left[a(y-y_0)\right]\!+\!\tan\left[a(y-y_0)\right]\right\}.
\end{eqnarray}

If the range of the fifth dimension is
$-\pi/2\le a(y-y_0)\le\pi/2,$ such a solution describes a single thick brane
located at $y_0$, where $a$ characterizes the width of the warp factor
$\Delta\sim 1/a$ in the case when the constant $b>0$. Thus, we have considered
just the positive definite part of the $\cos$ function in order to avoid possible sign
changes in the signature of the metric. This solution corresponds to the following
self--interacting potential
\begin{equation}
\label{intpot}
V=\frac{3a^2b}{8}\left\{\sec^2\left[a(y-y_0)\right]-2b\tan^2\left[a(y-y_0)\right]\right\},
\end{equation} which, after inverting (\ref{sol}), can be explicitly written
as a function of the scalar field:
\begin{equation}
\label{intpotfi} V(\phi)=\frac{3a^2b^2}{4}\left[1+\frac{1-2b}{b}
\cosh^2\left(\frac{2\phi}{\sqrt{3b}}\right)\right].
\end{equation}
A particularity of this solution is that it develops naked singularities at the boundaries of the
extra dimension, i.e., at $-\pi/2a$ and $\pi/2a$. Another interesting feature of this solution is
that the self--interacting potential (\ref{intpotfi}) is unbounded from below when $b>1/2$, a
situation that usually takes place in AdS supergravity without pathology \cite{dewolfe,townsend}.

We shall also consider another solution to our model originally derived in  \cite{gremm1} (similar 
solutions were constructed in \cite{csakietal} and \cite{kantietal3} in a slightly different context):
\begin{eqnarray}
{\bf Solution \ B)}\qquad \quad \label{sol2} A(y) = -b \ln\left[2 \cosh(c y) \right],\qquad \quad
\phi(y) = \sqrt{6b}\, {\rm arctan}
    \left[ \tanh\left( \frac{c y}{2}\right)\right],
\end{eqnarray}
where the domain of the extra dimension is infinite $-\infty<y<\infty$, with the corresponding potential
\begin{equation}
\label{intpotfi2} V(\phi) = \frac{3 c^2 b}{8} \left[
(1-4b)+(1+4b)\cos\left(2\sqrt{\frac{2}{3b}}\phi\right)\right].
\end{equation}
This solution can be interpreted as a thick AdS domain wall located at $y_0=0$ with two free
parameters: one for the width of the wall, given by $c,$ and another for the AdS curvature, which
is characterized by $bc$ \cite{gremm1}. The potential (\ref{intpotfi2}) represents an inverted potential and
also seems to be unstable from a simple point of view.

However, in \cite{Giovannini} it was proven that the scalar--tensor system (\ref{action}) that generates the 
braneworlds considered here is stable under scalar perturbations since the relevant dynamical equations for 
these scalar fluctuations can be expressed in a form inferred from supersymmetric quantum mechanics which 
guarantees that the spectrum is positive definite with no tachyonic modes with $m^2<0$. Below we shall go 
further by showing that for the considered solutions of our model, the corresponding analog quantum mechanical 
potentials are positive definite barriers along the fifth dimension and thus, the spectrum of scalar 
perturbations is delocalized from the corresponding braneworlds.

\subsection{Fluctuations of the metric and gravity localization}

Let us study the metric fluctuations $h_{mn}$ of (\ref{conflinee})
given by
\begin{equation}
\label{mfluct}
ds_5^2=e^{2A(y)}\left[\eta_{mn}+h_{mn}(x,y)\right]dx^m dx^n+dy^2.
\end{equation}
In the general case one must also consider the fluctuations of the
scalar field when treating fluctuations of the classical background
metric since they are coupled, however, following \cite{dewolfe}, we
shall just study the transverse traceless modes of the background
fluctuations $h_{mn}^T$ since they decouple from the scalar
perturbations.

We perform the coordinate transformation 
\begin{equation}
\label{coordtransf} 
dz=e^{-A}dy, 
\end{equation} 
in order to get a conformally flat metric. Thus, the equation for the transverse traceless modes 
of the metric fluctuations $h_{mn}^T$ reads (see, e.g., \cite{dewolfe}--\cite{csakietal} for details)
\begin{equation}
\label{eqttm}
\left(\partial_z^2+3A'\partial_z+\Box\right)h_{mn}^T=0,
\end{equation}
where $A'=dA/dz$. This equation supports a massless and normalizable
4D graviton given by $h_{mn}^T=C_{mn}e^{ipx}$, where $C_{mn}$ are
constant parameters and $p^2=m^2=0$.

In order to recast equation (\ref{eqttm}) into a Schr\"{o}dinger's
equation form, we adopt the ansatz
\begin{equation}h_{mn}^T=e^{ipx}e^{-3A/2}\Psi_{mn}(z) \label{Psi5}
\end{equation}
and get the equation
\begin{equation}
\label{schrodinger} [\partial_z^2-V_{QM}(z)+m^2]\Psi=0,
\end{equation}
where we have dropped the subscripts in $\Psi$, $m$ is the 4D mass of
the KK excitation modes, and the analog quantum mechanical
potential, which is completely defined by the curvature of the
manifold, reads
\begin{equation}
V_{QM}(z)=\frac{3}{2}\partial_z^2A+\frac{9}{4}(\partial_z A)^2.
\end{equation}
Thus, the warp factor determines the dynamics of the KK gravitational fluctuations as we shall see
further. Within the framework of the Quantum Mechanics, the spectrum of eigenvalues $m^2$
parameterizes the spectrum of graviton masses that a 4D observer standing at $z=0$ sees.

We shall further particularize the analysis of the physical properties of the graviton spectrum
predicted by Eq. (\ref{schrodinger}) for the above presented solutions A) and B) with the aim of
establishing whether localization of 4D gravity is feasible or not within our model. This study will also
allow us to determine what is the analytical structure of the massive graviton spectrum that is 
responsible for the corrections to the Newton's law.

\subsubsection{Localization of gravity for solution A)}

In the particular case when $b=2$ in solution A) we can invert the
coordinate transformation (\ref{coordtransf}) and explicitly express
the variable $y$ in terms of $z$ yielding the equality:\footnote{For this solution the case $b=2$ has also 
been studied in \cite{gremm1,csakietal}, but the corresponding Schr\"odinger 
equation has not been solved for arbitrary $m$. The point here is that, in general, it is really hard to express the 
coordinate $z$ as an invertible function of $y$.}
$$
\cos\left[a(y-y_0)\right]={\rm sech}(az),
$$
practically decompactifying the fifth dimension and sending the curvature singularities that were
present at $-\pi/2a$ and $\pi/2a$ to spatial infinity as in \cite{gremm2,bhrs}. However, it should
be pointed out that this coordinate transformation does not make the naked singularities harmless
in our model since they still are at a finite proper distance from the location of the $3$--brane.

In terms of $z$, the function $A(z)$ adopts the following form
\begin{equation}
A(z)=\ln{{\rm sech}(az)},
\label{AAz} 
\end{equation}
while the analog quantum mechanical potential becomes 
\begin{equation}
\label{potentialpes3e4} V_{QM}(z)=\frac{3a^2}{4}\left[3-5{\rm sech}^2(az)\right].
\end{equation}

First of all we note that by looking at the asymptotical behavior of the effective quantum
mechanical potential, it approaches the positive value $V_{QM}(\infty)=9a^2/4$, a fact that
guarantees the existence of a mass gap between the zero mode bound state and the first KK massive
mode \cite{csakietal}. This situation is quite generic when considering de Sitter 3--branes (see
\cite{ps,wang}), however, it is quite unusual when studying $3$--branes with Poincar\'e symmetry.
By performing the change of variable $u=az$ the Schr\"odinger equation can be transformed into the
standard form possessing a modified P\"oschl--Teller potential (see \cite{bhrs} and \cite{PTptl},
for instance)
\begin{equation}
\label{schrodingerPT} \left[-\partial_u^2-n(n+1){\rm sech}^2u\right]\Psi(u)=
\left[\frac{m^2}{a^2}-\frac{9}{4}\right]\Psi(u)=E\Psi(u),
\end{equation}
with $n=3/2$. Since the integer part of $n$ is equal to one, i.e. $[n]=1$, there exist two bound
states: the ground state $\Psi_0$ with energy $E_0$ and an excited state $\Psi_1$ with energy
$E_1$. In fact, the Schr\"odinger equation (\ref{schrodinger}) can be solved analytically and the
general solution can be expressed as a linear combination of associated Legendre functions of
first and second kind of degree $3/2$ and order $\mu=\sqrt{\frac{9}{4}-\frac{m^2}{a^2}}$:
\be
\Psi_m=k_1\ P_{3/2}^{\mu} \left[\tanh(az)\right]+k_2\ Q_{3/2}^{\mu}\left[\tanh(az)\right],
\label{Psi_m} \ee where $k_1$ and $k_2$ are integration constants. We already know that there are
two bound states in our potential; the ground state corresponds to the massless state $m=0$
($\mu=3/2$), and hence, possesses energy $E_0=-n^2=-9/4$, whereas the excited state has mass
$m=\sqrt{2}a$ ($\mu=1/2$) and energy $E_1=-(n-1)^2=-1/4$ .

When $\mu=1/2$ the series for our associated Legendre functions are finite according to
\cite{erdelyi}: \be P_{\nu}^{1/2}(z)= \sqrt{2\pi}~(z^{2}-1)^{-\frac{1}{4}}\left \{
\left[z+(z^{2}-1)^{\frac{1}{2}} \right ]^{\nu+\frac{1}{2}}+ \left[z+(z^{2}-1)^{\frac{1}{2}}
\right]^{-\nu-\frac{1}{2}} \right\}, \ee \be Q_{\nu}^{1/2}(z)=i
\sqrt{\frac{\pi}{2}}~(z^{2}-1)^{-\frac{1}{4}} \left[z+(z^{2}-1)^{\frac{1}{2}}
\right]^{-\nu-\frac{1}{2}}. \ee
A similar situation takes place for $\mu=3/2$ due to the following relations (see
\cite{gradshteyn})
\be P_{\nu}^{3/2}(z)=(z^2-1)^{-\frac{1}{2}}\left[(\nu-\frac{1}{2})z
P_{\nu}^{1/2}(z)-(\nu+\frac{1}{2})z P_{\nu-1}^{1/2}(z) \right], \ee \be
Q_{\nu}^{3/2}(z)=\left(z^{2}-1\right)^{-1/2} \left[(\nu-\frac{1}{2})zQ_{\nu}^{1/2}(z)-
(\nu+\frac{1}{2})Q_{\nu-1}^{1/2}(z)\right]. \ee
Thus, with the aid of these formulae one can obtain the following eigenfunction for the massless
mode ($\mu=3/2$) from the general solution (\ref{Psi_m}):
\be \Psi_0=C_0\ {\rm sech}^{3/2}(az), \label{psi0}\ee
which precisely coincides with the result $\Psi_0\sim e^{3A/2}$ predicted in \cite{csakietal} as an 
universal aspect of thick braneworlds.

In a similar way, the excited massive mode (with $\mu=1/2$) can be constructed from the general
solution (\ref{Psi_m}) and is given by the following expression
\be \Psi_1=C_1\ {\rm sinh}(az){\rm sech}^{3/2}(az). \label{psi1}\ee
Here $C_0$ and $C_1$ are normalization constants. Within the context of de Sitter 
braneworld models, in \cite{wang} it is erroneously stated
that when $\mu$ is real, or $m^2 < 9a^2/4$, both $P_{3/2}^{\mu}\left[\tanh(az)\right]$ and
$Q_{3/2}^{\mu}\left[\tanh(az)\right]$ are singular at $|z|\rightarrow\infty$ and hence, the
regularity conditions at spatial infinity for the metric perturbations exclude the case when $m^2
< 9a^2/4$. However, it is precisely the case when $\mu$ is real that gives rise to the analytical
expressions for the bound states of the spectrum (\ref{psi0}) and (\ref{psi1}). The point is that
while both $P_{3/2}^{\mu}\left[\tanh(az)\right]$ and $Q_{3/2}^{\mu}\left[\tanh(az)\right]$ are
singular at $|z|\rightarrow\infty$, their linear combination becomes asymptotically regular under
a suitable relation between the arbitrary constants $k_1$ and $k_2$, leading to explicit
analytical expressions for the bound states of the transverse traceless metric perturbations.

The function (\ref{psi0}) is the lowest energy eigenfunction of the Schr\"odinger equation
(\ref{schrodinger}) since it has no zeros, and can be interpreted as a massless 4D graviton with
no tachyonic instabilities from modes with $m^2<0$, whereas the function (\ref{psi1}) represents a
normalizable massive graviton localized on the brane. Thus, there exists a mass gap between these
two bound states, an important fact from the phenomenological point of view since it eliminates
the dangerous presence of arbitrarily light KK excitations, sending them to energies corresponding
to the scale of $a$ (see below). This remarkable situation usually takes place in braneworld models 
with an induced 4D de Sitter metric \cite{ps,wang}.

There is also a continuous spectrum of massive modes with $m\geq3a/2$ and, thus,
the order vanishes or becomes purely imaginary $\mu=i\rho$, where
$$\rho=\sqrt{(m/a)^2-9/4}.$$

This fact allows us to analytically study the behaviour of the
massive modes of the spectrum of KK excitations, a scarce phenomenon
when considering smooth brane configurations with Poincar\'e
symmetry since usually the coordinate transformation
(\ref{coordtransf}) cannot be explicitly inverted and then, the
Schr\"odinger equation cannot be integrated for $m^2\ne 0$ in the
language of the coordinate $z$ and one is forced to make use of
numerical analysis (see \cite{dewolfe,gremm1}, and \cite{bazeiaetal},
for instance).

Within the context of purely geometric braneworlds in a Weyl geometry, in \cite{bhrs} it was shown 
that when $2m>3a$ we have a continuous spectrum of eigenfunctions that
describe plane waves as they approach spatial infinity:
\begin{eqnarray}
\Psi^{\mu}_{\pm}(z)= C_{\pm}(\rho)P^{\pm i\rho }_{\frac{3}{2}}
\Bigl(\tanh\bigl(a\, z\bigr)\Bigr)\sim \frac{1}{\Gamma(1\mp
i\rho)}e^{\pm ia\rho z}. \label{Psipm}
\end{eqnarray}
These massive excitations are precisely the KK modes that give rise to small corrections to
Newton's law in 4D flat spacetime according to the formula derived in \cite{bs} (see also \cite{csakietal}).

Thus, the analytical study of the linear metric perturbations reflects the fact that there exists
a zero mode separated from a massive excitation and a continuous spectrum of massive KK modes by a
mass gap determined by the value of the constant $m=\sqrt{2}a$, a fact that, indeed, guarantees
the normalizability of the 4D graviton zero mode.

As pointed out above, if one considers localization of 4D gravity with a generic warped ansatz for
the metric, the presence of the mass gap in the graviton spectrum is responsible for the
development of naked singularities at the boundaries of the fifth dimension. This analysis was
made in terms of a simple, but generic relationship existing between the 5D curvature scalar, the
warp factor and the quantum mechanical potential which governs the dynamics of the linearized
transverse traceless modes of metric fluctuations in \cite{dago}.

We shall now show that these singularities can be made harmless by following the traditional point
of view which imposes unitary boundary conditions that guarantee 4D energy and momentum
conservation on the spectrum of KK modes as in \cite{GMZ,cohenkaplan} (see also
\cite{gremm2,dago}), providing a viable model from the physical point of view. The fact that the
manifold develops naked singularities at the boundaries does not matter if no conserved quantities
are allowed to leak out through the boundaries.

Let us recall how this comes about: The Poincar\'e isometries of metric (\ref{conflinee}) corresponds 
to 4D conservation laws. Thus,
by considering the translations generated by the Killing vectors $\xi^\mu_n=\delta^\mu_n$, where
$\mu$ is a 5D index, we can construct currents by contracting them with the stress tensor
\begin{equation}
J^\mu = T^{\mu\nu}\xi^{n}_\nu. \label{current}
\end{equation}
These currents satisfy a covariant conservation law of 4D energy and momentum
\begin{equation}
\frac{1}{\sqrt{g}}\partial_\mu\left(\sqrt{g} J^\mu\right) = 0. \label{conservationlaw}
\end{equation}
We further demand that the flux through the singular boundary of spacetime, i.e., along the
transverse direction, must vanish for all currents in order to ensure that these quantities are
conserved in the presence of a singularity:
\begin{equation}
\lim_{z\to\infty} \sqrt{g} J^z = \lim_{z\to\infty} \sqrt{g} g^{zz} \frac{1}{2} \partial_n
h^T_{pq}\partial_z h^T_{pq} = 0. \label{noflux}
\end{equation}

Thus, by recalling (\ref{Psi5}) and imposing the unitary boundary conditions to the asymptotic
form of $\Psi(z):$
\begin{equation}
\Psi(z)\sim A\sin(a\rho z)+B\cos(a\rho z),
\end{equation}
we get the following result
\begin{equation}
\lim_{z\to\infty} e^{3A/2}\Psi(z)\partial_z\left[e^{-3A/2}\Psi(z)\right]\sim
\Psi(z)\left[(3B-2A\rho)\cos(a\rho z)+(3A+2B\rho)\sin(a\rho z)\right] = 0, \label{noflux}
\end{equation}
implying that $A$ and $B$ must vanish identically, eliminating at once all the continuum massive
modes from the spectrum of KK excitations and rendering the naked singularities harmless. Thus,
the resulting unitary spectrum consists of just the two discrete modes since they do not generate
any flux into the singularities.

By summarizing, the graviton spectrum of the above considered solution A) contains a massless
graviton, one massive excited state with $m= \sqrt2 a$, and a continuum of modes with a mass gap
of size $m= 3a/2$. At very low energies, none of these massive modes can be excited, and an
observer at $z=0$ sees pure 4D gravity. At higher energies the massive state can be excited,
giving some corrections to Newton's law and, finally, at energies larger than the gap the whole
continuum of modes can be excited. Violations of unitarity occur only when modes that can travel
out to the singularities can be excited, i.e.~only at energies above the mass of the lightest
continuum mode\footnote{Below we shall see that these energies are of the order of the Planck mass 
for solution A) and are therefore experimentally unavailable for a 4D observer.}. When imposing 
unitary boundary conditions, the continuum sector of the spectrum is
projected out. Thus, the theory  remains unitary only for the discrete part of the spectrum
because the continuous massive modes die off rapidly enough as they approach towards the naked
singularity, rendering a physically viable model.

Alternatively, these naked singularities could be resolved either by lifting the 5D geometry to a
higher dimension or to string theory, since there exist examples reported in the literature where
5D naked singularities correspond to non--singular 10D spacetimes \cite{warner}.

\subsubsection{Localization of gravity for solution B)}

A similar analysis can be performed for the solution B) when
$b=1$: In this case, after applying the transformation
(\ref{coordtransf}) we get the following relation between the
coordinates $y$ and $z$:\footnote{In this case, the transformation
(\ref{coordtransf}) which defines $z$ in terms of $y$ involves 
integrals of the form $z \sim \int dy \cosh^b(cy)$, which are easy to perform for integer $b$. 
However, for even $b$ the inversion is never possible in closed form whereas for odd $b$ 
it requires solving a degree $b$ polynomial equation. The case $b = 3$ is also tractable, 
but the resulting equations are much more complicated.} 
\be 2\cosh(c(y-y_0))=\sqrt{4+c^2z^2}. \ee
Thus, the warp factor adopts the form
\be A(z)=-\ln{\sqrt{4+c^2z^2}}; 
\label{ABz}
\ee
the quantum mechanical potential reads
\be V_{QM}(z)=\frac{3c^2}{4}\frac{5c^2z^2-8}{(4+c^2z^2)^2} \label{VqmA}\ee
and has the form of a volcano potential with finite bottom which asymptotically vanishes,
indicating that there exists a single normalizable bound state with no mass gap in the graviton
spectrum of KK fluctuations for this particular case.

We shall now proceed to exactly solve the Schr\"odinger equation (\ref{schrodinger}) with the
potential (\ref{VqmA}) for the first time in order to analytically study the spectrum of massive KK modes for this
solution of the model. It turns out that it is easier to obtain the solution for this equation
with arbitrary $m$ if we go to the complex realm, get the solution there and then come back to the
original real variables. As mentioned above, these massive KK modes are responsible for the
corrections to Newton's law and we need them in order to analytically compute such corrections.

By applying the following transformations for the parameter $a=i\alpha,$ the fifth coordinate
$w=\alpha^2z^2/4,$ where the domain of the new coordinate is $0\le w<\infty,$ and the wave
function
\be \Psi=(1-w)^{-3/4}U(w), \label{wftransf} \ee
we recast the Schr\"odinger equation (\ref{schrodinger}) into the Ince's limit of the confluent
Heun equation, or equivalently, of the generalized spheroidal wave equation (see \cite{figueiredo,BF}), given by
\begin{eqnarray}
\label{IncelimitBF} w(w-w_{0})\frac{d^{2}U}{dw^{2}}+(B_{1}+B_{2}w)
\frac{dU}{dw}+ \left[B_{3}+q(w-w_{0})\right]U=0,\qquad q\neq0,
\end{eqnarray}
with $w_0=1,$ $B_1=-1/2,$ $B_2=-1,$ $B_3=0$ and
$q=m^2/\alpha^2;$ thus, our equation adopts the form
\begin{eqnarray}
\label{Incelimit}
w(w-1)\frac{d^{2}U}{dw^{2}}-\left(\frac{1}{2}+w\right)
\frac{dU}{dw}+ \frac{m^2}{\alpha^2}\left(w-1\right)U=0,
\end{eqnarray}
where $w=0$ and $w=w_0=1$ are regular singularities and infinity is an irregular one according to
the classification given in \cite{slavyanovlay}. Here we should point out that the Ince's limit of
the generalized spheroidal wave equation requires solutions (with arbitrary mass $m>0$) behaving
at infinity as the so-called subnormal Thom\'e solutions quoted in \cite{olver,slavyanovlay}:
\begin{eqnarray}\label{thome}
\lim_{w\rightarrow  \infty}U(w)\sim e^{\pm
2i\sqrt{qw}}w^{(1/4)-(B_{2}/2)}=e^{\pm
2i\frac{m}{|\alpha|}\sqrt{w}}w^{3/4},
\end{eqnarray}
in contrast to solutions with asymptotic normal Thom\'e behaviour of a confluent and double
confluent Heun equations \cite{ronveaux,BF}. This particular behaviour of the solutions
corresponds to the description of plane waves at spatial infinity in the extra dimension by the
wave function (\ref{wftransf}) as we shall see further. In fact, these massive modes generate
small corrections to Newton's law coming from the extra dimension and having analytic expressions
for them is extremely useful when explicitly computing such corrections.

It turns out that B. Figueiredo has constructed several solutions to this type of differential
equation in \cite{figueiredo}. A complete solution consists of a pair of two--sided infinite
series defined in different domains. One branch of the solution in each pair is given by a
two--sided infinite series of hypergeometric functions and converges for any finite value of the
independent variable $w$, while the other branch is given in terms of a two--sided infinite series
of modified Bessel functions and converges for $|w|>|w_{0}|$, where $w_{0}=1$ denotes a regular
singularity. As mentioned above, these solutions possess an arbitrary phase parameter $\nu$ and
are two--sided in the sense that the summation index $n$ runs from $-\infty$ to $\infty$. However,
if there is an arbitrary parameter in the equation, one can use the phase parameter $\nu$ in order
to truncate the series and get one--sided solutions with $n\geq 0$ (see \cite{figueiredo,BF} for
details and illustrative examples). Since we do not have such an arbitrary parameter at our
disposal, here we shall quote one of these two--sided solutions, the so--called first pair in
\cite{figueiredo}:
\begin{eqnarray}
\label{solU}
\begin{array}{l}
U^{0}= \displaystyle
\sum_{n=-\infty}^{\infty}b_{n}
F\left(\frac{B_2}{2}-n-\nu-1,n+\nu+\frac{B_{2}}{2};B_{2}
+\frac{B_{1}}{w_{0}};1-\frac{w}{w_{0}}\right), \vspace{5mm}\\
U^{\infty} =w^{\frac{1-B_{2}}{2}}\,\displaystyle
\sum_{n=-\infty}^{\infty}b_{n}
K_{2n+2\nu+1}\left(\pm2i\sqrt{q w}\right),
\end{array}
\end{eqnarray}
where the superscript `zero' means that the series converges in any
finite part of the complex plane, while the superscript `infinity'
indicates convergence for $\vert w\vert>\vert w_{0}\vert$ and the
coefficients $b_{n}$ obey three--term recurrence relations
\begin{eqnarray}
\label{r1a}
\alpha_{n}b_{n+1}+\beta_{n}b_{n}+\gamma_{n}b_{n-1}=0,
\end{eqnarray}
and can be calculated with the aid of the following quantities
\begin{eqnarray}
\begin{array}{l}
\alpha_{n} = \frac{qw_{0}\ \left(n+\nu+2-\frac{B_2}{2}\right) \left(n+\nu+1
-\frac{B_2}{2}-\frac{B_{1}}{w_{0}}\right)}
{\left(n+\nu+1\right)\left(n+\nu+\frac{3}{2}\right)},
\vspace{.2cm} \\
\beta_{n} = 4B_{3}-2q w_{0}+4 \left(n+\nu+1-\frac{B_{2}}{2}\right)
\left(n+\nu+\frac{B_{2}}{2}\right) -\frac{2q
w_{0}\left(\frac{B_{2}}{2}-1\right)
\left(\frac{B_{2}}{2}+\frac{B_{1}}{w_{0}}\right)}
{\left(n+\nu\right)
\left(n+\nu+1\right)},
\vspace{.3cm} \\
\gamma_{n} = \frac{q w_{0}\
\left(n+\nu+\frac{B_{2}}{2}-1\right)
\left(n+\nu+\frac{B_{2}}{2}+\frac{B_{1}}{w_{0}}\right)}
{\left(n+\nu-\frac{1}{2}\right)
\left(n+\nu\right)},
\end{array}
\end{eqnarray}
Thus, in our case, the pair of solutions adopts the following form:
\begin{eqnarray}
U^{0}= \displaystyle \sum_{n=-\infty}^{\infty}b_{n}\
F\left(-n-\nu-\frac{3}{2},n+\nu-\frac{1}{2};-\frac{3}{2};1-w\right),
\nonumber
\end{eqnarray}
\begin{eqnarray}
U^{\infty} =w\displaystyle \sum_{n=-\infty}^{\infty}b_{n}\
K_{2n+2\nu+1}\left(\pm2i\frac{m}{|\alpha|}\sqrt{w}\right), \qquad \mbox{with} \label{solnU}
\end{eqnarray}
\begin{eqnarray}
\begin{array}{l}
\alpha_{n}\!=\!\frac{m^2\left(n+\nu+\frac{5}{2}\right)
\left(n+\nu+2\right)}
{\alpha^2\left(n+\nu+1\right)\left(n+\nu+\frac{3}{2}\right)}, \\
\beta_{n}\!=\!-\!2\frac{m^2}{\alpha^2}\!+\!4\left(n\!+\!\nu\!+\!\frac{3}{2}\right)
\left(n\!+\!\nu\!-\!\frac{1}{2}\right)\!-\!\frac{3m^2}{\alpha^2\left(n+\nu\right)
\left(n+\nu+1\right)},
\\
\gamma_{n}=\frac{m^2\left(n+\nu-\frac{3}{2}\right)\left(n+\nu-1\right)}
{\alpha^2\left(n+\nu\right)\left(n+\nu-\frac{1}{2}\right)}.
\end{array}\nonumber
\end{eqnarray}

Computing an explicit solution like (\ref{solnU}) involves the calculation of the coefficients
$b_n,$ which obey (\ref{r1a}) and possess a dominant and a minimal solution (see
\cite{figueiredo}, for instance). What we are actually doing here is a little bit trickier: We are 
producing two minimal solutions for (\ref{r1a}). One of them is minimal in the forward sense 
--and is generated with a backward recursion-- while the other is minimal in the backward 
sense --and is generated with a forward recursion. These two solutions are then pasted in 
the origin and normalized accordingly. The minimal solution thus constructed guarantees 
the convergence of the two--sided infinite series (\ref{solU}) and must be chosen as an 
original physical solution to the Ince's limit of the confluent Heun equation (\ref{IncelimitBF}) 
and, hence, to the relevant Schr\"odinger equation (\ref{schrodinger}) of our problem.

The zero mode can easily be computed by looking at the solution
(\ref{solnU}) in the massless case $m=0\ (q=0).$ Thus, the parameters
$\alpha_n$ and $\gamma_n$ vanish and the remaining parameters
$\beta_n$ are all non--zero, a fact which implies that all the
coefficients $b_n =0.$ On the other hand, Eq. (\ref{Incelimit})
tells us that when $m=0$ the function $U(w)$ can be a constant since
the integration is defined up to an arbitrary constant. Thus, the
only bound state of the system is given by the following
eigenfunction 
\be \Psi_0=\frac{k_0}{\left(4+c^2z^2\right)^{3/4}},
\qquad\qquad k_0=\mbox{const.} 
\label{psizero}
\ee 
and corresponds to the
normalizable 4D graviton, free of tachyonic instabilities, as
expected from the solution of the massless Schr\"odinger equation
$\Psi_0\sim e^{3A/2}$.

It is worth noticing that since the solution (\ref{solnU}) presents a subnormal Thom\'e behaviour
at spatial infinity according to (\ref{thome}), then, when we come back to the wave function
$\Psi$ through the transformation (\ref{wftransf}) and return to the language of the parameter $a$
and the original coordinate $z,$ the pair of solutions (\ref{solnU}) to the Schr\"odinger equation
(\ref{schrodinger}) transforms into
\begin{eqnarray}
\Psi^{0}= \left(1+\frac{a^2z^2}{4}\right)^{-3/4}\displaystyle \sum_{n=-\infty}^{\infty}b_{n}\
F\left(-n-\nu-\frac{3}{2},n+\nu-\frac{1}{2};-\frac{3}{2};1+\frac{a^2z^2}{4}\right), \nonumber
\end{eqnarray}
\begin{eqnarray}
\Psi^{\infty} =-\frac{a^2z^2}{4}\left(1+\frac{a^2z^2}{4}\right)^{-3/4}\displaystyle
\sum_{n=-\infty}^{\infty}b_{n}\ K_{2n+2\nu+1}\left(\pm imz\right) \label{solnPsiz}
\end{eqnarray}
with the same coefficients $\alpha_n$, $\beta_n$ and $\gamma_n$ which obey the three term
recurrence relations (\ref{r1a}). Moreover, the asymptotic behaviour of these solutions, as
physically expected, corresponds to plane waves:
\begin{eqnarray}\label{planewaves2}
\lim_{z\rightarrow  \infty}\Psi(z)\sim e^{\pm imz}z^{3/2}\
(4+a^2z^2)^{-3/4}\sim e^{\pm imz}.
\end{eqnarray}
We shall make use of these massive modes in Subsection $2.4$ where we shall explicitly compute the
corrections to Newton's law of gravitation for the solution B) of the model.

It is quite remarkable that the solution (\ref{psizero})--(\ref{solnPsiz}) represents an original application of the 
InceÕs limit of the confluent Heun equation (or generalized spheroidal wave equation) within the framework 
of thick braneworld models generated by gravity coupled to a bulk scalar field.

\subsection{Stability of the system under scalar perturbations}

We shall now briefly review the stability analysis of the considered braneworld under linear perturbations of 
the scalar sector following the results reported in \cite{Giovannini}, where the master equations for this 
scalar--tensor system were derived. Then we shall go further by elucidating the structure of the mass spectra 
of scalar fluctuations from the form of the analog quantum mechanical potentials corresponding to the 
relevant Schr\"odinger--like equations. 

Following \cite{Giovannini} we consider the perturbed metric for the scalar sector of fluctuations written in 
the longitudinal gauge: 
\begin{equation}
ds^2 = e^{2A(z)}\left[\left(1+2\psi(x^{\mu},z)\right)\eta_{\mu\nu}dx^{\mu}dx^{\nu}+\left(1+2\xi(x^{\mu},z)\right)dz^2\right],
\end{equation}
along with the background scalar field perturbations $\phi(x^{\mu},z)=\phi(z)+\chi(x^{\mu},z)$, where $\psi$, 
$\xi$ and $\chi$ are considered small fluctuations. It turns out that the system of coupled differential 
equations for all of these fluctuations can be reduced to a pair of master equations of Schr\"odinger--like form. 
However, there is just one independent degree of freedom, i.e. only one scalar physical mode, after taking 
into account the constraints $\xi=2\psi$ and $\chi=-\frac{6}{\phi'}\left(\psi'+2A'\psi\right)$.  

Thus, by introducing the following rescaled scalar perturbation $Y=\frac{e^{3A/2}}{\phi'}\psi$ we can recast 
its relevant dynamical equation into the Schr\"odinger form 
\begin{equation}
Y'' - g \left(g^{-1}\right)'' Y = m_{Y}^2Y ,
\label{Psieqn}
\end{equation}
where we have implemented a convenient separation of 4d and 5d variables which involves the definition of the 4d mass 
$m_{Y}$ and we have introduced the following function 
\begin{equation}
g = \frac{e^{3A/2}\phi'}{A'} ,
\label{g}
\end{equation}
which cleverly parameterizes the analog quantum mechanical potential
\begin{equation}
U_{Y} = g \left(g^{-1}\right)''. 
\label{VQMPsi}
\end{equation}
The same equation holds for the rescaled scalar fluctuation $\xi$ in view of the constraints mentioned 
above.  Parallelly, the dynamical equation for the new scalar perturbation $X=e^{3A/2}\chi-g\psi$ 
also adopts the Schr\"odinger--like form
\begin{equation}
X'' - g^{-1}g'' X = m_X^2 X,
\label{Xeqn}
\end{equation}
after defining its 4d mass $m_X$. We see that the analog quantum mechanical potential now reads
\begin{equation}
U_{X} = g^{-1}g'',
\label{VQMX}
\end{equation}
and is related to the potential (\ref{VQMPsi}) by the following discrete symmetry $g\rightarrow g^{-1}$. 
It turns out that both of these potentials can respectively be expressed as 
\begin{equation}
U_{Y} = {\cal J}_{Y}^2 - {\cal J}_{Y}', \qquad\qquad U_X = {\cal J}_X^2 + {\cal J}_X', 
\label{pots}
\end{equation}
in the terms of the so--called superpotentials ${\cal J}_{i} = \frac{g'}{g}$, where $i=Y,X$.
These quantities were inferred from supersymmetric quantum mechanics and are very useful since in 
their language the relevant Schr\"odinger equations for $Y$ and $X$ adopt the form
\begin{equation}
{\cal Q}^{\dagger} {\cal Q} Y = m^2_{Y} Y, \qquad\qquad  {\cal Q} {\cal Q}^{\dagger} X = m^2_X X,
\end{equation}
where the operators ${\cal Q}^{\dagger}$ and ${\cal Q}$ are defined as follows:
\begin{equation}
{\cal Q}^{\dagger} = \biggl( - \frac{d}{dw} + {\cal J} \biggr), \qquad\qquad
{\cal Q} = \biggl( \frac{d}{dw} + {\cal J} \biggr).
\label{operators}
\end{equation}

Therefore, since the Schr\"odinger equations for $Y$ and $X$ can be expressed in this form, 
it follows that in the corresponding spectra of scalar fluctuations there are no tachyonic modes 
with negative mass and, thus, our braneworld configuration is stable under linear scalar perturbations.

In order to establish whether or not there are localized perturbations in the scalar sector we must study 
the behaviour of the analog quantum mechanical potentials (\ref{VQMPsi}) and (\ref{VQMX}) of the 
corresponding Schr\"odinger equations (\ref{Psieqn}) and (\ref{Xeqn}) since this analysis must be 
performed for each particular solution separately. 

It turns out that for solution A) the warp factor is given by (\ref{AAz}) and the scalar field adopts the form 
$\phi_A=\sqrt{6}az$, hence, we get the following function $g_A\sim {\rm sech}^{3/2}(az)/\tanh(az)$. 
On the other side, for solution B) the warp factor is (\ref{ABz}) and the scalar field reads 
$\phi_B=\sqrt{6}\,{\rm arctan}\left[ {\rm tanh}\left(\frac{1}{2}{\rm arccosh}\sqrt{1 + c^2 z^2/4}\right)\right]$, 
yielding the following function $g_B\sim z^{-1}\left(4+c^2 z^2\right)^{-3/4}$. It is straightforward to compute 
the analog quantum mechanical potentials $U_i(z)$ for both solutions A) and B) and see that the four quantities 
represent positive definite potential barriers along the fifth dimension. 

This fact implies that the scalar modes are not localized on the brane within the framework of our stable 
braneworld model for both solutions A) and B).

\subsection{Mass hierarchy problem}

In order to be able to aim at the mass hierarchy problem within the present thick brane scenario,
one has to add a (in principle thin, positive tension) probe brane some distance away from the
location of the thick (Planck) brane\footnotemark\footnotetext{Since there is no precise meaning
to the term ``thick brane location", one could think of it as the location $z=0$ ($y=y_0$) where
the graviton wave--function is peaked.} -- where 4D gravity is bound -- in the $z$--direction. It
is supposed, besides, that the SM particles are trapped in the (thin, positive tension) probe
brane. In order to get the 4D effective Planck scale, we replace the Minkowski metric by a 4D
metric $\overline{g}_{mn}$ in equation (\ref{mfluct}), leading to an effective 4D action after
integrating over $z$ in (\ref{action}). As customary we look at the curvature term from which one
can derive the scale of the gravitational interactions
\begin{equation}
\label{actioneffective} S_{eff}\supset 2M_{\ast}^3 \int d^4x
\int_{-\infty}^{\infty} dz\sqrt{|\overline{g}|}e^{3A}\overline{R},
\end{equation}
where we performed the coordinate transformation $dy=e^{A}dz$, and $\overline{R}$ is the 4D Ricci
scalar. The effective Planck mass can be straightforwardly calculated:
\begin{equation}
M_{pl}^2=2M_{\ast}^3\int_{-\infty}^{\infty}dz
e^{3A}=\frac{M_{\ast}^3}{\beta},
\end{equation}
which is finite for both of our solutions (for the case A) we get $\beta=a/\pi$, while for the
solution B) we have $\beta=c/2)$; if we take $M_{\ast}\sim\beta\sim M_{pl}$, then the zero mode
$\Psi_0$ is coupled correctly to generate 4D (Newtonian, in particular) gravity.

Suppose the TeV probe brane were located at some definite position
$z_0$ in the extra--space, then the SM particles see the metric:
\begin{equation}
g_{ab}^{SM}=e^{2A(z_0)}\overline{g}_{ab}={\rm
sech}^2(\beta z_0)\overline{g}_{ab},
\end{equation} for
both of our solutions. Hence, following the approach of
Reference \cite{rs}, the physical mass scales are set by the
symmetry--breaking scale
\begin{equation} m={\rm
sech}(\beta z_0)\;m_0.
\end{equation}
To produce TeV physical mass scales from fundamental Planck mass
parameters, we need for our solutions $${\rm sech}(\beta z_0)\approx
\frac{TeV}{M_{pl}},
$$ i.e.,
$\beta z_0\approx \ln 2+16\ln 10\approx 38$. Recalling that the thickness
of the Planck (thick) brane is $\Delta\sim 1/\beta\approx
M_{pl}^{-1}$, then the probe (TeV) brane has to be placed 40 times
the thickness of the Planck brane away from the origin $z=0$, where
the graviton wave-function is peaked. This means that even
experiments probing energies far bigger than TeV scale would not be
able to resolve the above mentioned separation.

There is another aspect of the problem that needs to be discussed
for the case A). That standard 4D (Newtonian) gravity is reproduced
on the thick (Planck) brane, is clear from the fact that there is a
mass gap of energy $m\sim a\sim M_{pl}$ between the stable (ground
state) graviton $\Psi_0$ and the normalizable massive graviton
$\Psi_1$, that is also bound to the Planck brane. The continuous
modes have masses bigger than $\Psi_1$ and, since these are
delocalized, the corresponding amplitudes are suppressed at the
origin with respect to the amplitudes $\Psi_0$ and $\Psi_1$.
However, since the probe brane where the SM particles live is
located away from the origin, one could think that the massive
continuous modes could play an important role in modifying
gravitational interactions at the TeV brane, as long as the ratio of
the corresponding amplitudes to $\Psi_0$ (and $\Psi_1$) grows as one
recedes from $z=0$ in the extra-space (an effect related with the
localization of $\Psi_0$ and $\Psi_1$ on the thick brane and with
the delocalization of the KK--continuum). This is true (even if we
need only a separation $z_0\approx 40/M_{pl}$ from the Planck brane
to generate the correct hierarchy), but recall that the masses of
the corresponding continuous modes are of the order of the Planck
mass and bigger. Consequently, a 4D observer placed at the TeV brane
would not be able to detect those modes. As the probe brane position
recedes from the origin the energies accessible to probe brane
observers decreases, as well as the possibility to detect massive
modes of the KK--continuum. The consequence is that, as long as
correct Newtonian gravity is achieved at the Planck brane, gravity
on a probe brane located at any position in the extra--space will be
Newtonian as well, confirming the result obtained in \cite{lr}.

As for the solution B) in which there is no mass gap between the 4D
normalizable graviton bound state and the continuum of massive
modes, a situation similar to that of the Randall--Sundrum and
Lykken--Randall models takes place, since the 4D graviton is localized
at the thick brane and is ``protected" by the barriers of
the quantum mechanical potential from the continuum of delocalized
massive modes which are suppressed at $z_0$ due to an analog tunneling
effect. On the other hand, since the thickness of the brane is inversely
proportional to the Planck mass $\Delta\sim 1/c\approx M_{pl}^{-1}$ as
stated above, it is extremely small to be resolved by 4D observers
located at the TeV probe brane in this case as well. Thus, this
circumstance, together with the fact that plane waves coming from
infinity decay subasymptotically (since they must tunnel trough the
potential barriers to reach the $z_0$ position, where the massless
graviton is localized), leads to a recovery of 4D gravity very similar
to that of the Randall--Sundrum and Lykken--Randall models as we shall
see from the analytic computation of small corrections to Newton's law.
It is worth noticing that for this special solution to our model the 5D
Ricci scalar is regular everywhere as can be seen from its expression:
\begin{equation}
R=-8e^{-2A}\left(A_{zz}+\frac{3}{2}A_z^2\right)=4a^2\frac{8-5a^2z^2}{4+a^2z^2},
\label{R5}
\end{equation}
leading to a 5D manifold which is completely free of naked singularities, in contrast with the
previous case.

Here it is worthwhile to mention some words about the stability problem of the brane position
$z_0$ along the fifth dimension. For both studied cases, the introduction of the probe TeV brane
away from the Planck brane in order to achieve the desired warping, and hence, the desired
hierarchy, gives rise to a new fine tuning on the probe brane position, and therefore, to the need
of stabilizing this brane separation. It turns out that the stabilization of this brane separation
is achieved through the Goldberger--Wise mechanism by associating to it a canonical scalar field
with interaction terms on both branes that models the radius of the fifth dimension when one
neglects the back--reaction of the probe brane \cite{GW}. Subsequently, in \cite{dewolfe} the
back--reaction of the probe brane was incorporated and it was shown that when considering the full
back--reacted system, the brane separation remains stable when modeled by a canonical scalar field
with a family of quartic self--interaction potentials. Furthermore, these authors also
obtained the desired warping from the Planck scale to the TeV one, thus resolving the fine tuning
problem of the Higgs mass in a stable braneworld scenario\footnote{This problem contrasts with the
situation that arises when studying the localization of scalar field fluctuations on branes.
Notwithstanding, scalar field fluctuations are localized on the brane only when one neglects their
gravitational back--reaction (i.e., when one assumes that the scalar field makes a small
contribution to the energy of the bulk and does not modify its geometry, so that the background
solutions remain valid in the presence of the bulk scalar field), on the contrary, as far as one
considers their back--reaction, scalar fluctuations necessarily delocalize from the brane.}.

\subsection{KK corrections to Newton's law}

Let us consider the thin brane limit $\beta\to\infty$, where we can
locate two test bodies at the center of the brane in the transverse
direction and compute the corrections to Newton's law in 4D flat
spacetime coming from the fifth dimension that are generated by the
massive gravitons and can be expressed as in \cite{bs}:
\begin{eqnarray}
U(r)\sim\frac{M_1M_2}{r}\biggl(G_4+M_{\ast}^{-3} \sum_i e^{-m_ir}
\vert\Psi_i(z_0)\vert^2+M_{\ast}^{-3}\int_{m_0}^{\infty}dm\,e^{-mr}
\vert{\Psi^{\mu(m)}(z_0)}\vert^2\biggr)\nonumber\\
=\frac{M_1M_2}{r}\biggl(G_4+\Delta G_4\biggr), \label{U}
\end{eqnarray}
where the thin brane is located at $z=z_0$ and represents the physical universe, $G_4$ is the 4D
gravitational coupling, $\Psi_i$ represents the wave functions of the discrete excited states with
mass $m_i$ (if any), and $\Psi^{\mu(m)}$ denotes the continuous eigenfunctions.

Within the thin brane limit of solution A), since the $\Psi_1(z_0)$ wave function is odd, it does
not contribute to the corrections, moreover, the integral of (\ref{U}) is dominated by the small -
$\rho$ region and it can be well--approximated by expanding the prefactor of the exponential at
$\rho =0$. In this way one obtains the following expression for the corrections $\Delta G_4$ to 4D
Newton's law  (see \cite{bhrs} for details of a similar result within a Weyl integrable model)
\begin{eqnarray}
\Delta G_4 \sim&M_{\ast}^{-3}
\frac{1}{\left\vert\Gamma(-\frac{1}{4})
\Gamma(\frac{7}{4})\right\vert^2}
\frac{1}{ar}e^{-\frac{3}{2}ar}\left(1+ O(\frac{1}{ar})\right).
\label{approxDeltaGN}
\end{eqnarray}
It is worth noticing that away from the thin brane limit, the correction will involve the massive
bound state and the form of Newton's law will depend on the precise location of the two test
bodies along the extra dimension. Recall that this result is valid if small violations of unitarity 
are allowed in the model. Thus, when momentaneously ignoring the presence of naked singularities at 
the boundaries of the manifold of  
the brane solution A), the corrections to Newton's law are exponentially suppressed as one recedes 
from the brane along the fifth dimension, where $a$ is of order of the Planck mass, making these 
corrections particularly small. Moreover, it is usually believed that the numerical coefficient 
entering the corrections to Newton's law $\Delta G_4$ is of order unity. An interesting fact this 
explicit calculation shows is that such a coefficient is of order $\left\vert\Gamma(-\frac{1}{4})
\Gamma(\frac{7}{4})\right\vert^{-2}\sim 10^{-2}$ making the exponentially suppressed corrections
to Newton's law a hundred times smaller, a physical consequence which is due to the existence of 
the mass gap in solution A).

In the case of solution B) there are no discrete massive modes and thus, there is no second term
in (\ref{U}). After evaluating the massive wave functions (\ref{solnPsiz}) at $z_0=0,$ computing the
corresponding series and substituting their expression in (\ref{U}), where integration must be
performed  with respect to $dm/a$ as pointed out in \cite{bs}, we get the following corrections to Newton's law in
this particular case:
\begin{eqnarray}
\Delta G_4 \sim&M_{\ast}^{-3} \frac{1}{2\pi a^2 r^2} \left(1+
O\left(\frac{1}{(ar)^2}\right)\right),
\label{approxDeltaGN2}
\end{eqnarray}
a result that coincides with the RS correction up to a factor of $1/2\pi$.

Corrections to Newton's law were obtained for brane models with a non--minimally coupled bulk
scalar field in \cite{farakosetal}.

\section{Concluding Remarks}

We have considered particular scalar thick brane generalizations of the RS model in which 4D
gravity is localized at a certain point of the fifth dimension. These field configurations do not
restrict the 5D spacetime to be an orbifold geometry by smoothing out the singularities at the
brane position, and avoid the introduction of thin branes in the action by hand.

When studying linear metric perturbations, for the two particular solutions A) and B) we obtain 
analytic expressions for the lowest energy eigenfunction which represents a single bound state 
that can be interpreted as a stable 4D graviton free of tachyonic modes with $m^2<0$.

For the first solution A) there is also an excited KK massive mode separated from the massless
graviton state by a mass gap determined by the scale of the constant $a$, inversely proportional
to the thickness of the brane. The appearance of the mass gap eliminates the dangerous low--energy
KK graviton excitations from the theory under consideration (in the considered special case), a
quite relevant aspect from the phenomenological point of view. The gap is a consequence of the
fact that the quantum analog potential of the Schr\"odinger--like equation, obtained after
linearizing 5D gravity, asymptotes to a positive constant, unlike most part of related models,
where the potential asymptotes to zero, as in case B), where immediately after the massless
bound state there exists a continuum of massive modes (see, for instance,
\cite{dewolfe}--\cite{bazeiaetal} and \cite{thickbranes}).

We get as well a continuous spectrum of massive KK modes that starts at $m=3a/2$ for solution
A), and at $m>0$ for solution B), and turn into continuum plane waves as they approach spatial
infinity.

The massive modes of KK excitations are explicitly given in terms of associated Legendre functions
of first and second kind for solution A), and in terms of two--sided infinite series of
hypergeometric functions and modified Bessel functions for the case B), allowing for analytical
computations of corrections to Newton's law along the lines of \cite{csakietal} and \cite{bhrs},
in contraposition to implemented numerical approaches \cite{gremm1,bazeiaetal} because of the lack
of exact solutions for these KK modes.

We have also shown that for both solutions A) and B) the considered scalar--tensor braneworld model 
is stable under linear scalar perturbations with no massive modes localized on the brane even though 
at first sight the corresponding self--interaction potentials of the scalar field seemed to be 
problematic.

One of the main achievements of the present paper is that, following the procedure of
\cite{rs,lr}, a resolution of the mass hierarchy is given by placing a positive--tension probe
brane -- where the SM particles are trapped -- some distance away from the thick (Planck) brane in
the extra--space in both considered cases.

As already discussed, for solution A) there is a mass gap of order of the Planck mass scale, that
provides a clue to solving the mass hierarchy while recovering Newtonian gravity on the TeV brane
in a clear and simple way. This solution to our model develops naked singularities at the
boundaries of the 5D manifold \cite{gremm2,bhrs} as a consequence of the existence of the mass gap
in accordance to \cite{dago}. However, we impose unitary boundary conditions on the graviton
spectrum, ensuring that no supposedly conserved quantities disappear into the singularity. This
approach was successfully used in \cite{cohenkaplan} in order to render a specific naked
singularity harmless. Thus, by analyzing the resulting spectrum of the KK massive modes after
applying the unitary boundary conditions, we see that they remove the continuum sector of the
spectrum for the solution A) of the model discussed here. It is straightforward to check that the
two discrete modes do not generate any flux into the singularity, so that the unitary spectrum
consists of just these two bound states. Since the massive bound state is odd with respect to the
fifth dimension, it does not contribute to the correction to Newton's law and we recover just 4D
gravity on the brane after imposing unitary boundary conditions on the solution A) of the model.
In the case of solution B) there is no mass gap in the spectrum of KK excitations, however, we
show that the mass hierarchy problem can be solved in a similar way with positive branes and 
that Newtonian gravity is
also recovered. A good peculiarity of this solution consists in rendering a smooth 5D curvature
scalar free of naked singularities along the extra dimension, in contrast with solution A).

Finally, small corrections to 4D Newton's law were analytically computed for both of our solutions
in the thin brane limit, showing that this braneworld model can be suitable to address the above
mentioned problems. Moreover, for the case A) the mass gap in the graviton spectrum leads to an
exponential suppression of the corrections to Newton's inverse square law, favored by an extra 
suppression of order $10^{-2}$ due to the numerical factor that appears in them and that we were
able to explicitly compute. It is worth mentioning that the parameter $a$ is of Planck mass order,
rendering very small corrections when allowing for small violations of unitarity (when the problem
of naked singularities is left aside). For the case B), these corrections obey a power law behaviour which 
coincides with the Randall--Sundrum result up to a numerical coefficient equal to $1/2\pi$. These 
are explicit analytical examples of the qualitative computation of corrections to Newton's law 
outlined in \cite{csakietal} under general assumptions that were calculated thanks to the exact 
solutions we constructed to both the Legendre equation and to the Ince's limit of the confluent 
Heun equation, for cases A) and B), respectively. It should be stressed that the CHE in the Ince 
limit finds a novel physical application for the first time within the framework of thick 
braneworld models.

\section*{Acknowledgements}

Two of the authors (AHA and IQ) are really grateful to C. Germani, R. Maartens, D.
Malag\'on--Morej\'on, and S.L. Parameswaran for fruitful and illuminating discussions while this
investigation was carried out. AHA and KK also thank B. Figueiredo for useful correspondence and
comments. AHA is grateful to the staff of the ICF, UNAM and MCTP, UNACH for hospitality. 
This research was supported by grants CIC--UMSNH, CONACYT 60060--J, COECYT, Instituto Avanzado 
de Cosmolog\'\i a (IAC), the MES of Cuba, PAPIIT--UNAM, No. IN103413-3, {\it Teor\'ias de Kaluza-Klein, inflaci\'on y 
perturbaciones gravitacionales,} and by the research grant 89298 (2013) granted from the 
Research Committee of the Aristotle University of Thessaloniki. NBC acknowledges a postdoctoral grant from 
DGAPA--UNAM. KK is grateful to the Institute of Physics and Mathematics (IFM) of the University of Michoacan (UMSNH) for hospitality and acknowledges support from a postdoctoral scholarship during 2011 from the Research Committee of the Aristotle University of Thessaloniki. NBC, AHA and UN are grateful to SNI for support, while IQ was supported by 
``Programa PRO--SNI, Universidad de Guadalajara"â under grant No. 146912.

\end{document}